\documentclass[a4paper,english]{article}
\usepackage[T1]{fontenc}
\usepackage[latin1]{inputenc}
\usepackage{subfigure}
\usepackage{amsmath}
\usepackage{graphicx}
\usepackage{amssymb}

\makeatletter

\providecommand{\LyX}{L\kern-.1667em\lower.25em\hbox{Y}\kern-.125emX\@}

\def\one{\mbox{\rm 1}\hskip-2.8pt \mbox{\rm l}}
 \newcommand{\lyxaddress}[1]{
   \par {\raggedright #1
   \vspace{1.4em}
   \noindent\par}
 }

\usepackage{babel}
\makeatother
\begin{document}

\title{On bialgebras associated with paths and essential paths on ADE graphs}

\author{R. Coquereaux${}^1$ \ and
             A. O. Garc\'{\i}a${}^2$\thanks{~Email: coque@cpt.univ-mrs.fr, garcia@iwr.fzk.de
             \newline \hspace*{0.3cm}$^{**}$ Present address.
             \newline
             \newline \hspace*{0.3cm}$^{}$ CPT-2004/P.074}}

\maketitle

\lyxaddress{${}^1${\small Centre de Physique Th\'eorique, Case 907, Luminy, 13009,
Marseille, France.}}
\lyxaddress{${}^2${\small Max-Planck-Institut f\"ur Physik, F\"ohringer Ring 6, 80805 M\"unchen, Germany,
and Forschungszentrum Karlsruhe, Postfach 3640, 76021 Karlsruhe, Germany$^{**}$.}}

\begin{abstract}
We define a graded multiplication on the vector space of
essential paths on a graph $G$ (a tree) and show that it is associative.
 In most interesting applications,  this tree  is an ADE Dynkin diagram.
The vector space of length preserving endomorphisms of essential paths 
has a grading obtained from the length of paths and
possesses several interesting bialgebra structures. 
One of these, the Double Triangle Algebra (DTA) of A. Ocneanu, 
is a particular kind of quantum groupoid (a weak Hopf algebra) and was studied 
elsewhere; its coproduct gives a filtrated convolution 
product on the dual vector space. Another bialgebra structure 
is obtained by replacing this filtered convolution product by a graded associative product.
It can be obtained from the former by projection on a subspace of 
maximal grade, but it is interesting to define it directly, without using 
the DTA. What is obtained is a weak bialgebra, not a weak Hopf algebra.

\end{abstract}

\section*{Introduction\label{sec:introduction}}

Paths of a given length $n$ between two vertices $a,b$ of a Dynkin diagram 
(extended or not) can be interpreted in terms of classical or quantum $SU(2)$ 
intertwiners between representations $a \otimes \tau^n$ and $b$, where $\tau$ 
is the fundamental (spin $1/2$). In a similar way, essential paths are 
associated with (classical or quantum) morphisms between $a \otimes \tau_n$ 
and $b$, where $\tau_n$ denotes an irreducible representation.

We consider the graded vector space of essential paths (defined by A. Ocneanu 
for quite general graphs) and its algebra of grade-preserving endomorphisms.
The corresponding associative product called {\sl composition product\/} is 
denoted $\circ$.

We first show that the space of essential paths carries an associative algebra 
structure (denoted $\bullet$) compatible with its natural grading. Its definition 
involves the usual concatenation product of paths, but the situation  is 
not trivial since the concatenation product of two essential paths is usually 
not essential.
This is actually our main result, and it seems to be new (the existing 
literature is more concerned with the algebra structures that can be defined 
at the level of the {\em graded} tensor square of this vector space). 

Using this ``improved'' concatenation product between essential paths, one 
can then define --besides the composition product-- two other interesting algebra 
structures on the algebra of grade-preserving endomorphisms.
One of these algebra structures (denoted $\star$) is associated with  a filtrated 
convolution product and gives rise to a weak Hopf algebra structure: this 
is the Double Triangle Algebra (DTA) introduced by A. Ocneanu in 
\cite{Ocneanu:paths}. It was studied elsewhere (see \cite{PetZub:Oc}, 
\cite{Gil:thesis}, \cite{CoTr-DTE}).
Another algebra structure, that we call the graded convolution product or 
simply\footnote{Both $\bullet$ and $\star$ can be understood as convolution 
products. Given two elements of grades $p$ and $q$, the composition product 
is trivial unless $p=q$, the graded product gives an element of grade
$p+q$ whereas the ``filtered product'' can be decomposed along vector subspaces 
of all grades $p+q, p+q-2, p+q-4,\ldots $} {\sl graded product}, and again 
denote by the symbol $\bullet$, can be obtained from the former product by 
projection on its component of highest degree. However, it is possible and 
useful to study it directly without making any reference to its filtered 
relative. This is what we do.

Both products $\bullet$ and $\star$ are compatible with the composition of
endomorphisms $\circ$. Compatibility here means that the associated coproducts 
are algebra homomorphisms with respect to the composition product. The 
 use of a particular scalar product allows one to study these 
three product structures on the same underlying vector space (the diagonal graded
tensor square of the space of essential paths).
The bialgebra associated with the pair $(\circ, \star)$ is known to be a particular kind 
of quantum groupoid. However, in this paper we are interested in the bialgebra 
associated with the pair $(\circ, \bullet)$, and we show that it has a weaker 
structure: it is a weak bialgebra but not a weak Hopf (bi)-algebra. 

The whole theory should apply when the diagrams that we consider (usually 
ADE Dynkin diagrams) are replaced by members of higher Coxeter-Dynkin systems \cite{DiFZub,Ocneanu:Bariloche}: the vector space spanned by the vertices 
of the chosen diagram is, in particular, a module over the graph algebra 
associated with a Weyl alcove of $SU(N)$ at some level ---such generalised 
${\cal A}$ diagrams are indeed obtained by truncation of the Weyl chamber 
of $SU(N)$. These systems admit also orbifolds ---${\cal D}$ diagrams--- and 
exceptionnals. In the higher cases, the grading does not refer to the positive 
integer that measures the length of a path, but to a particular Young diagram. 
Therefore the grading is defined with respect to a more general monoid (actually 
an integral positive cone), and the adjective ``filtrated'' should be 
understood accordingly.

Our paper is organized as follows. In the first section we consider the
vector space of all paths on a graph, and show that it is a non-unital 
bialgebra. In section~\ref{sec:the-EssPaths-algebra} we restrict our 
attention to the subspace of essential paths and show that we need to 
introduce a new associative multiplication, $\bullet$, involving an 
appropriate projection operator, in order to insure stability.
This vector space of essential paths is an algebra, but not a bialgebra.
In the third section we show that the graded algebra of endomorphisms of 
essential paths can be endowed with a new product compatible with the 
grading, for which this space is a weak bialgebra. The non trivial 
condition insuring compatibility of the coproduct with the multiplication 
of endomorphisms is exemplified at the end of section~3, in the case of 
the graph $E_6$. The equation expressing this general condition is 
obtained in appendix~A, and the general proof showing that such a 
compatibility condition always holds in our situation is given in section~\ref{sec:weak_bialgebra_condition_proof}. In the fifth section we 
illustrate, in the case of the graph $A_2$, the fact that the two bialgebra 
structures respectively associated with the products $(\circ, \star)$ and 
$(\circ, \bullet)$ differ. Appendix~A is quite general: we consider an 
arbitrary algebra $A$ endowed with a scalar product and we show that 
although its endomorphism algebra can be given a coalgebra structure, 
some non trivial relation has to be satisfied in order for this space 
to be a bialgebra ---the coproduct on $End(A)$ should be an homomorphism.
We also study what happens when the algebra $A$ is graded and when we replace 
$End(A)$ by a graded diagonal sum of endomorphisms.

\section{The space of paths on a graph\label{sec:paths_on_graph}}

Take a connected and simply connected graph $G$. For the time being,
we do not assume any other extra requirements. At a later stage we will
take $G$ to be a tree, and, even more precisely, a Dynkin diagram of type ADE. 
For instance, a possible graph could be $G=E_{6}$.

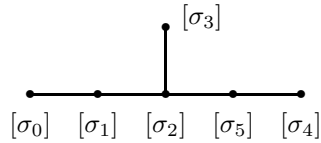
\begin{figure}[htb]
\unitlength 0.6mm
\begin{center}
\begin{picture}(95,35)
\thinlines 
\multiput(25,10)(15,0){5}{\circle*{2}}
\put(55,25){\circle*{2}}
\thicklines
\put(25,10){\line(1,0){60}}
\put(55,10){\line(0,1){15}}
\put(25,3){\makebox(0,0){$[\sigma_0]$}}
\put(40,3){\makebox(0,0){$[\sigma_1]$}}
\put(55,3){\makebox(0,0){$[\sigma_2]$}}
\put(70,3){\makebox(0,0){$[\sigma_5]$}}
\put(85,3){\makebox(0,0){$[\sigma_4]$}}
\put(63,27){\makebox(0,0){$[\sigma_3]$}}
\end{picture}
\label{graphE6}
\end{center}
\end{figure}

Consider the set of elementary paths on $G$. These are just
ordered lists of \emph{neighboring} points $a_{i}$ (or edges $\xi _{k}$
joining two neighboring points) of the graph, \[
[a_{0},a_{1},a_{2},\cdots ,a_{L-1},a_{L}]\qquad a_{i}\in G\]
This is clearly a path of length $L$, starting at $a_{0}$ and ending
at $a_{L}$. Build a vector space, called $Paths$, by simply considering
formal linear combinations over $\mathbb{C}$ of elementary paths.
Now define the product of elementary paths by concatenation, ie,
by joining the matching endpoints of the two paths (say, of lengths 
$L$ and $K$) one after the other,
\begin{eqnarray*}
[a_{0},a_{1},\cdots ,a_{L}]\, [b_{0},b_{1},\cdots ,b_{K}] & = & \left\{
\begin{array}{l}
[a_{0},a_{1},\cdots ,a_{L},b_{1},\cdots ,b_{K}] \quad \textrm{if} \ a_{L}=b_{0}\\
0 \qquad \qquad \qquad \qquad \qquad \qquad \textrm{otherwise}
\end{array}\right.
\end{eqnarray*}
Such an operation creates another elementary path of length $L+K$. This product extends
by linearity to the whole vector space, and is associative (this is
trivial to see). Moreover, the resulting algebra is \emph{graded}
by the length of the paths.

Consider additionally the zero-length paths $[a_{0}]$, there will
be one such for each point $a_{0}$ of the graph. If the graph is finite,
the sum over all points of the graph of the corresponding zero-length
paths will be a (left and right) unit for this algebra,\[
\mathbf{1}=\sum _{a_{0}\in G}\: [a_{0}]\]
Therefore $Paths$ is a graded associative algebra with unit.

We could also define a coalgebra structure on this space, introducing
a coproduct that would be group-like for all elementary paths $p$:

\[
\Delta p=p\otimes p\]
and extending it by linearity. It is straightforward to see that it
is coassociative and that it is an algebra homomorphism, 
$\Delta (pp')=\Delta p\: \Delta p'$. Additionally, the (linear) operation
\[\epsilon (p)=1\qquad \textrm{for all elementary} \; p \]
is a counit for $\Delta $.

The above defined unit is not compatible with the coproduct
($\Delta \mathbf{1}\neq \mathbf{1}\otimes \mathbf{1}$). 
$Paths$ is therefore a (non-unital) bialgebra.  It
is infinite dimensional even if the graph $G$ is finite, as paths
can be made arbitrarily long by backtracking on $G$.

However, as we shall see in the next section, the space $\mathcal{E}$ 
of essential paths that we consider in this paper is only a vector 
subspace but not a subalgebra of $Paths$.
For this reason, a different approach will be required.

\section{The algebra $\mathcal{E}$ of essential paths\label{sec:the-EssPaths-algebra}}

\subsection{Essential paths on a graph\label{sec:ess-paths-on-graph}}

We will now briefly introduce  essential paths on the given graph $G$.
Consider first the adjacency matrix of the graph, and call $\beta $
its maximal eigenvalue. Also call $\vec{\mu }=(\mu _{0}=1,\mu _{1},\cdots ,\mu _{N})$
the corresponding eigenvector, normalized such that the entry $\mu _{0}$
associated to a distinguished point $0$ of $G$ is equal to $1$ (this component
is minimal).
 $\vec{\mu }$ is called the Perron-Frobenius eigenvector, and all its components
are strictly positive. The next step is to introduce the linear
operators \[
C_{k}:Paths\longmapsto Paths\qquad \qquad k=1,2,3,\cdots \]
which act on elementary paths as follows: 
on a path of length $L \leq k$, $C_k$ gives zero, otherwise ($L>k$) its action is given  by,
\begin{eqnarray*}
& & C_{k}\left([a_{0},a_{1},\cdots ,a_{k-1},a_{k},a_{k+1},a_{k+2},\cdots ,a_{L}]\right) \\
& & \qquad \qquad \qquad = \: \delta _{a_{k-1},a_{k+1}}\: \sqrt{\frac{\mu _{a_{k}}}{\mu _{a_{k-1}}}}
 \quad [a_{0},a_{1},\cdots ,a_{k-1},a_{k+2},\cdots ,a_{L}]
\end{eqnarray*}
These operators obviously preserve the end-points of the paths they
act upon, and shorten their length by $2$ units ---removing a backtrack in the path
at position $k$, if any, and giving $0$ otherwise.

The {\em essential paths} are defined as those elements of $Paths$ annihilated
by all the $C_{k}$'s. They constitute, of course, a vector subspace%
\footnote{If the graph is a Dynkin diagram of type ADE then $\beta < 2 $ 
and $\mathcal{E}$ is finite dimensional, as there are essential paths up 
to a certain length only, namely from $0$ to $\kappa - 1$, where 
$\kappa$ is the Coxeter number of the diagram defined by 
$\beta = 2\, cos (\pi/\kappa)$.} 
$\mathcal{E}\subset Paths$:\[
\mathcal{E}=\left\{ p\in Paths\, \diagup \, C_{k}p=0\; \forall k\right\} \]
We will use $\mathcal{E}_{l}$ to denote the subspace of essential
paths of length $l$, and $\mathcal{E}(a\stackrel{l}{\longrightarrow }b)$
if we want to further restrict the set to those paths with definite
starting point $a$ and ending point $b$. 

On the whole $Paths$ there is a natural scalar product, defined on
elementary paths $p,p'$ by \[
\left\langle p,p'\right\rangle =\delta _{p,p'}\qquad \qquad (p,p'\; \textrm{elementary})\]
 and consequently also an orthogonal projector \[
P:Paths\longmapsto \mathcal{E}\]
As paths with different lengths or end-points are orthogonal, $P$
can be decomposed as a sum of projectors on each subspace,
\begin{eqnarray*}
P & = & \sum _{\begin{array}{c}
 a,b\in G\\
 l\in \mathbb{N}\end{array}
}\, P_{ab}^{l}\\
P_{ab}^{l} & : & Paths(a\stackrel{l}{\longrightarrow }b)\longmapsto \mathcal{E}(a\stackrel{l}{\longrightarrow }b)
\end{eqnarray*}

We had on $Paths$ an algebra structure, but actually $\mathcal{E}$
is only a vector subspace and not a subalgebra of $Paths$. Therefore,
a new product has to be found on $\mathcal{E}$ if we want to endow it
with an algebra structure. The simplest one (it must also be somehow
related to the one on $Paths$!), is:
\begin{equation}
e\bullet e'\equiv P(ee')\label{eq:product-on-EP}
\end{equation}
where $e,e'$ are essential paths, $P$ is the above orthogonal projector
and the product $ee'$ is the concatenation product in $Paths$. We shall 
prove below the associativity property and find a unit element for this product.

\subsubsection{The grading of $\mathcal{E}$}

As we did with $Paths$, the space of essential paths can be graded
by the length of the paths,\[
\mathcal{E}=\bigoplus _{l\in \mathbb{N}}\mathcal{E}_{l}\]
The product $\bullet $ is clearly compatible with this grading because
$\left\langle \: ,\, \right\rangle $ is null for paths with different
lengths, hence the projector $P$ also preserves the length. For this
reason, we shall call it the \underline{graded} product on $\mathcal{E}$.
As stated in the Introduction, it is possible to define also a
\underline{filtered} product on the same space (which is called $\times$
in \cite{Coque-Cocoyoc}), such that $p \times p'$ can be decomposed 
on paths of lengths smaller or equal to $length(p) + length(p')$.
Moreover, the graded product $\bullet$ could be obtained from the filtered 
one by restriction to the component of highest length, although this approach will 
not be followed here.

\subsubsection{Example of essential paths on $E_{6}$}

The space $\mathcal{E}(E_{6})$ can be constructed using the above
definitions, and is of dimension $156$. More precisely, the
dimensions of the graded components are $(6,10,14,18,20,20,20,18,14,10,6)$.
For instance, the subspace $\mathcal{E}_{2}$ of paths of length $2$ has dimension $14$.
It is composed of a subspace corresponding to paths with different endpoints
plus a $4$-dimensional subspace of paths with coinciding ends. Inside the
latter there is a 2-dimensional subspace of paths which start and end at the
point $2$, which is generated by:
\begin{eqnarray*}
\mathcal{E}(2\stackrel{2}{\longrightarrow }2) & = & \left\{ \frac{1}{N_{1}}\left([2,3,2]-\sqrt{\frac{\mu _{3}}{\mu _{1}}}\, [2,1,2]\right)\right.,\\
 &  & \left.\frac{1}{N_{2}}\left([2,5,2]-\frac{\sqrt{\mu _{3}\mu _{5}}}{\mu _{1}+\mu _{3}}\, [2,3,2]-\frac{\sqrt{\mu _{1}\mu _{5}}}{\mu _{1}+\mu _{3}}\, [2,1,2]\right)\right\} \\
 & = & \left\{ \frac{1}{N_{1}}\left([2,3,2]-\sqrt{-1+\sqrt{3}}\, [2,1,2]\right)\right.\\
 &  & \left.\frac{1}{N_{2}}\left([2,5,2]-\frac{1}{\sqrt{3}}\, \sqrt{-1+\sqrt{3}}\, [2,3,2]-\frac{1}{\sqrt{3}}\, [2,1,2]\right)\right\} 
\end{eqnarray*}
These paths are orthogonal, and can be normalized with an appropriate
choice of the coefficients $N_{i}$.

\subsection{Associativity}

The product $\bullet $ in $\mathcal{E}$ is associative. In fact,
we will prove a stronger condition for the operator $P$, which implies
associativity of $\bullet $:\begin{equation}
P(P(p_{1})P(p_{2}))=P(p_{1}p_{2})\qquad \textrm{for}\; \textrm{any}\quad p_{i}\in Paths\label{eq:assoc_condition_on_P}\end{equation}
To see this take $\; e,e',e''\in \mathcal{E}\; $ then\begin{eqnarray*}
(e\bullet e')\bullet e'' & = & P(P(ee')\, e'')=P(P(ee')\, P(e''))=P((ee')\, e'')\\
 & = & P(e\, (e'e''))=P(P(e)\, P(e'e''))=P(e\, P(e'e''))\\
 & = & e\bullet (e'\bullet e'')
\end{eqnarray*}
The condition (\ref{eq:assoc_condition_on_P}) may also be rewritten
in the completely equivalent way\begin{eqnarray*}
P(P(p_{1})P(p_{2}))=P(p_{1}p_{2}) & \Longleftrightarrow  & P(P(p_{1})P(p_{2})-p_{1}p_{2})=0\\
 & \Longleftrightarrow  & I\equiv \left\langle e,p_{1}p_{2}-P(p_{1})P(p_{2})\right\rangle =0\qquad \textrm{for}\; \textrm{all}\quad e\in \mathcal{E}
\end{eqnarray*}
Now we have to show that $\; I=0\; $ for any $\; p_{i}\in Paths$:

\begin{itemize}
\item If $\; p_{1},p_{2}\in \mathcal{E}\subset Paths\; $ then $\quad P(p_{i})=p_{i}$
$\quad \Longrightarrow \quad $ $p_{1}p_{2}-P(p_{1})P(p_{2})=0$ $\quad \Longrightarrow \quad $
$I=0$.
\item If $\; p_{1}\equiv e_{1}\in \mathcal{E}\; $ but $\; p_{2}\in Paths\; $
then

\[
I=\left\langle e,e_{1}(p_{2}-P(p_{2}))\right\rangle =\left\langle e,e_{1}n\right\rangle \]
 Here $\; n\equiv p_{2}-P(p_{2})\in \mathcal{E}^{\perp }\; $ is orthogonal
to $\mathcal{E}$. 

Without loss of  generality, we may assume that the paths involved
in $I$ have well defined end-points and length (it is enough to show
associativity for such paths, then associativity for linear combinations
of those follows immediately): \begin{eqnarray*}
p_{1}=e_{1}=e_{1}(a\stackrel{l_{1}}{\longrightarrow }b) &  & \\
p_{2}=p_{2}(b'\stackrel{l_{2}}{\longrightarrow }c) & \; \Rightarrow \;  & n=n(b'\stackrel{l_{2}}{\longrightarrow }c)
\end{eqnarray*}

To get a non-trivial scalar product in $I$ we must also take $b'=b$
and

\[
e=e(a\stackrel{l_{1}+l_{2}}{\longrightarrow }c)\]

As it will be proven in subsection \ref{sub:decomposition-of-EP}, such
an essential path $e$ can always be decomposed as: \[
e=\sum _{\begin{array}{c}
 v\in G\\
 i_{v}\end{array}
}e_{i_{v}}'(a\stackrel{l_{1}}{\longrightarrow }v)\: e_{i_{v}}''(v\stackrel{l_{2}}{\longrightarrow }c)\]
 where the sum runs over all intermediate points $v$ appearing in
$e$ after $l_{1}$ steps, and possibly several $e_{i_{v}}'$, $e_{i_{v}}''$
for each $v$. Essentiality of $e$ and linear independence of paths
of different end-points imply that all the $e_{i_{v}}'$ and $e_{i_{v}}''$
are also essential. But now it is easy to see that \begin{eqnarray*}
I=\left\langle e,e_{1}n\right\rangle  & = & \left\langle \sum _{v,i_{v}}\, e_{i_{v}}'(a\stackrel{l_{1}}{\longrightarrow }v)\: e_{i_{v}}''(v\stackrel{l_{2}}{\longrightarrow }c)\; ,\; e_{1}(a\stackrel{l_{1}}{\longrightarrow }b)\, n(b\stackrel{l_{2}}{\longrightarrow }c)\right\rangle \\
 & = & \sum _{i_{b}}\left\langle e_{i_{b}}'(a\stackrel{l_{1}}{\longrightarrow }b)\: e_{i_{b}}''(b\stackrel{l_{2}}{\longrightarrow }c)\; ,\; e_{1}n\right\rangle \\
 & = & \sum _{i_{b}}\left\langle e_{i_{b}}'\: ,\: e_{1}\right\rangle \left\langle e_{i_{b}}''\: ,\: n\right\rangle 
\end{eqnarray*}
Therefore we get $I=0$ because $\; n\perp \mathcal{E}$, so $\; \left\langle e_{i_{b}}''\: ,\: n\right\rangle =0$.

\item If both $\; p_{1},p_{2}\in Paths\; $ then $\; p_{i}=e_{i}+n_{i}\; $
with $\; P(p_{i})=e_{i}$

Therefore\begin{eqnarray*}
I & = & \left\langle e,(e_{1}+n_{1})(e_{2}+n_{2})-e_{1}e_{2}\right\rangle =\left\langle e,e_{1}n_{2}+n_{1}e_{2}+n_{1}n_{2}\right\rangle \\
 & = & 0
\end{eqnarray*}
due to the previous case.

\end{itemize}

\subsection{Unit element}

The algebra $\mathcal{E}$ is unital, and the unit element is clearly
the same as the one in $Paths$, explicitly given by\begin{equation}
\mathbf{1}_{\mathcal{E}}=\sum _{v\in G}e(v\stackrel{0}{\longrightarrow }v)\label{eq:unit_on_EP}\end{equation}
where the sum extends over all the points of the graph, and the essential
paths $e(v\stackrel{0}{\longrightarrow }v)$ are obviously nothing
more than the trivial paths $e(v\stackrel{0}{\longrightarrow }v)\equiv [v]$.

Concluding this section, we  emphasize that $\mathcal{E}$
is not only a vector space but also an associative algebra. Moreover,
it is endowed with a (canonical) scalar product obtained by restriction
from the one on $Paths$. It has therefore also a coalgebra structure%
\footnote{Identify elements with their duals, and map the product to the dual
coproduct.%
},  which is not a priori very interesting since the comultiplication will
not be an algebra homomorphism in general.  The coproduct that we had defined 
for $Paths$ does not work either (the compatibility property with the product 
does not hold) since the product itself was modified. Therefore, contrary 
to $Paths$, the vector space $\mathcal{E}$ endowed with the
graded multiplication $\bullet$ does not have a bialgebra structure.

\section{The weak-$*$-bialgebra $End_{\#}(\mathcal{E})$}

We have already shown in section \ref{sec:the-EssPaths-algebra} that
the space $\mathcal{E}$ of essential paths constitutes a graded unital
associative algebra. Applying the general construction of Appendix~A
(see in particular Eq.~(\ref{eq:homo_condition_graded})) to the particular 
case of the graded algebra $A=\mathcal{E}$, we show now that a corresponding 
weak bialgebra structure on the space of its graded endomorphisms does exist.
Moreover, we shall see that it has a compatible star operation. 

We remind again the reader that the product $\bullet$ that we consider now on
$End_{\#}(\mathcal{E})$ is graded but that it is possible to construct another 
product (called $\star$) on the same vector space, which is filtered rather 
than graded. Moreover, the structure corresponding to the pair 
($\circ$, $\star$) is a weak Hopf algebra. This other construction is 
not studied in the present paper. What we obtain here instead, is a weak 
bialgebra structure for the pair ($\circ$, $\bullet$).

\subsection{Product and coproduct}

$\mathcal{E}$ being  a graded algebra, its endomorphisms can also
be graded. We therefore consider the space $\mathcal{B}$ of length
preserving endomorphisms on $\mathcal{E}$, namely

\begin{eqnarray*}
\mathcal{B}\: \equiv \: End_{\#}(\mathcal{E}) & = & \bigoplus _{n}End(\mathcal{E}_{n})\\
 & \stackrel{iso}{\simeq } & \bigoplus _{n}\mathcal{E}_{n}\otimes \mathcal{E}_{n}^{*}
\end{eqnarray*}
As discussed in section \ref{sub:graded-End(A)}, we now consider  the convolution product $\bullet $ on the
space of these endomorphisms. Recalling (\ref{eq:convolution_product})
we see that it is determined by the product on the algebra
$\mathcal{E}$, which we had also denoted by $\bullet $, meaning
concatenation of paths plus re-projection on the essential subspace.
Explicitly, on monomials we have\[
(e_{i}\otimes e^{j})\bullet (e_{k}\otimes e^{l})=e_{i}\bullet e_{k}\otimes e^{j}\bullet e^{l}\]
We also take the coproduct (\ref{eq:graded_composition_coproduct}),
which reads\[
\Delta \left(e_{i}\otimes e^{j}\right)=\sum _{I}\left(e_{i}\otimes e^{(n)I}\right)\otimes \left(e_{I}^{(n)}\otimes e^{j}\right)\qquad \qquad \textrm{whenever}\quad e_{i}\in \mathcal{E}_{n}\: ,\; e^{j}\in \mathcal{E}_{n}^{*}\]
but remark that the compatibility condition (\ref{eq:homo_condition_graded})
still remains to be verified. This will be done for a general graph
later (see section \ref{sec:weak_bialgebra_condition_proof}), but for any 
given graph it is interesting to explicitly check
equation (\ref{eq:homo_condition_graded}); we illustrate this
below in the case of the graph $E_{6}$.

\subsubsection{Case $E_{6}$}

As an example, we look at the highly non-trivial case of the graph $E_{6}$.
We shall consider normalized essential paths of length $4$ on $E_6$ and 
show how they appear in the $\bullet$ products of essential paths of length 
$2$ (this is just one possibility among others, of course). We have a natural 
coproduct on the dual but also, using the chosen scalar product, a coproduct 
on the same space of essential paths. Hence, we may use the previous 
calculation to find the expression of the coproduct $D$ of a particular 
essential path of length $4$ ---at least, that part which decomposes on the 
tensor products of essential paths of length $2$. Finally, we check that the 
compatibility condition described by Eq.~(\ref{eq:homo_condition_graded})
is satisfied, so that we can be sure, in advance, that the corresponding 
graded endomorphism algebra is indeed a week bialgebra.

The subspace $\mathcal{E}(2\stackrel{4}{\longrightarrow}2)$ of essential paths of length $4$
is $3$-dimensional and generated by the orthonormalized essential
paths $e_{1}(2\stackrel{4}{\longrightarrow }2)$, $e_{2}(2\stackrel{4}{\longrightarrow }2)$
and $e_{3}(2\stackrel{4}{\longrightarrow }2)$. With our convention
for choosing the basis the first two read explicitly, up to a normalization
factor,\begin{eqnarray*}
e_{1}(2\stackrel{4}{\longrightarrow }2) & \propto  & \frac{1}{\sqrt{2}}\, \sqrt{1+\sqrt{3}}\, \left([2,3,2,1,2]-[2,3,2,5,2]\right)\\
 &  & -\left([2,5,2,1,2]-[2,5,2,5,2]\right)-\sqrt{1+\sqrt{3}}\: [2,5,4,5,2]\\
e_{2}(2\stackrel{4}{\longrightarrow }2) & \propto  & \sqrt{1+\sqrt{3}}\: [2,1,0,1,2]-\left([2,1,2,1,2]-[2,1,2,5,2]\right)\\
 &  & +\frac{\sqrt{3}}{2}\, \sqrt{-1+\sqrt{3}}\, \left([2,3,2,1,2]-[2,3,2,5,2]\right)\\
 &  & +\frac{1}{2}\left(-1+\sqrt{3}\right)\, \left([2,5,2,1,2]-[2,5,2,5,2]\right)\\
 &  & +\frac{1}{\sqrt{2}}\, \sqrt{-1+\sqrt{3}}\: [2,5,4,5,2]
\end{eqnarray*}
 The generator $e_{2}(2\stackrel{4}{\longrightarrow }2)$ appears
as a component in some products of essential paths of length $2$,
namely in those products involving paths which have the point $2$
as one of the endpoints. These are:\begin{eqnarray*}
e(0\stackrel{2}{\longrightarrow }2) & = & [0,1,2]\\
e(2\stackrel{2}{\longrightarrow }0) & = & [2,1,0]\\
 &  & \\
e(2\stackrel{2}{\longrightarrow }4) & = & [2,5,4]\\
e(4\stackrel{2}{\longrightarrow }2) & = & [4,5,2]\\
 &  & \\
e_{1}(2\stackrel{2}{\longrightarrow }2) & \propto  & -\sqrt{-1+\sqrt{3}}\: [2,1,2]+[2,3,2]\\
e_{2}(2\stackrel{2}{\longrightarrow }2) & \propto  & -[2,1,2]-\sqrt{-1+\sqrt{3}}\: [2,3,2]+\sqrt{3}\: [2,5,2]
\end{eqnarray*}
The non-trivial products having a contribution in the direction $e_{2}(2\stackrel{4}{\longrightarrow }2)$
are

\begin{eqnarray*}
e(2\stackrel{2}{\longrightarrow }0)\bullet e(0\stackrel{2}{\longrightarrow }2) & = & \sqrt{1-\frac{1}{\sqrt{3}}}\: e_{2}(2\stackrel{4}{\longrightarrow }2)+\cdots \\
e_{1}(2\stackrel{2}{\longrightarrow }2)\bullet e_{1}(2\stackrel{2}{\longrightarrow }2) & = & -\frac{1}{\sqrt{6\sqrt{3}}}\: e_{2}(2\stackrel{4}{\longrightarrow }2)+\cdots \\
e_{1}(2\stackrel{2}{\longrightarrow }2)\bullet e_{2}(2\stackrel{2}{\longrightarrow }2) & = & -\frac{1}{3}\sqrt{\frac{3}{2}+\sqrt{3}}\: e_{2}(2\stackrel{4}{\longrightarrow }2)+\cdots \\
e_{2}(2\stackrel{2}{\longrightarrow }2)\bullet e_{1}(2\stackrel{2}{\longrightarrow }2) & = & -\sqrt{-\frac{4}{3}+\frac{7}{3\sqrt{3}}}\: e_{2}(2\stackrel{4}{\longrightarrow }2)+\cdots \\
e_{2}(2\stackrel{2}{\longrightarrow }2)\bullet e_{2}(2\stackrel{2}{\longrightarrow }2) & = & -\frac{1}{3}\sqrt{-3+2\sqrt{3}}\: e_{2}(2\stackrel{4}{\longrightarrow }2)+\cdots \\
e(2\stackrel{2}{\longrightarrow }4)\bullet e(4\stackrel{2}{\longrightarrow }2) & = & \sqrt{\frac{3}{2}-\frac{5}{2\sqrt{3}}}\: e_{2}(2\stackrel{4}{\longrightarrow }2)+\cdots 
\end{eqnarray*}
The factors preceding $e_{2}(2\stackrel{4}{\longrightarrow }2)$
in the above formulas are the coefficients $m_{ij}^{k}$ which enter
(\ref{eq:m_A_basis}) and (\ref{eq:graded_sum_mm_condition}). The
sum of the squares of the above six coefficients equals $1$, and this shows, 
in a particular example, how condition Eq.~(\ref{eq:homo_condition_graded})
can be checked (remember that it should be satisfied for each definite grading
of the coproducts of all elements).

Using (\ref{eq:D_A_basis}) we may also write 
$De_{2}(2\stackrel{4}{\longrightarrow }2)$ as
\begin{eqnarray*}
&& \sqrt{1-\frac{1}{\sqrt{3}}}\: e(2\stackrel{2}{\longrightarrow }0)\otimes e(0\stackrel{2}{\longrightarrow }2)
-\frac{1}{\sqrt{6\sqrt{3}}}\: e_{1}(2\stackrel{2}{\longrightarrow }2)\otimes e_{1}(2\stackrel{2}{\longrightarrow }2)\\
&& -\frac{1}{3}\sqrt{\frac{3}{2}+\sqrt{3}}\: e_{1}(2\stackrel{2}{\longrightarrow }2)\otimes e_{2}(2\stackrel{2}{\longrightarrow }2)
-\sqrt{-\frac{4}{3}+\frac{7}{3\sqrt{3}}}\: e_{2}(2\stackrel{2}{\longrightarrow }2)\otimes e_{1}(2\stackrel{2}{\longrightarrow }2)\\
&& -\frac{1}{3}\sqrt{-3+2\sqrt{3}}\: e_{2}(2\stackrel{2}{\longrightarrow }2)\otimes e_{2}(2\stackrel{2}{\longrightarrow }2)
+\sqrt{\frac{3}{2}-\frac{5}{2\sqrt{3}}}\: e(2\stackrel{2}{\longrightarrow }4)\otimes e(4\stackrel{2}{\longrightarrow }2)\\
&& +\cdots 
\end{eqnarray*}
where the  missing terms include tensor products of paths of
lengths $(3,1)$, $(1,3)$, $(0,4)$, and $(4,0)$. The last two
are clearly $[2]\otimes e_{2}(2\stackrel{4}{\longrightarrow }2)+e_{2}(2\stackrel{4}{\longrightarrow }2)\otimes [2]$.

As we will show explicitly%
\footnote{This also follows immediately from (\ref{eq:homo_condition_graded})
once this requirement is checked%
} in section \ref{sec:weak_bialgebra_condition_proof}, this also means
that the path $e_{2}(2\stackrel{4}{\longrightarrow }2)$ itself can
be decomposed as \begin{eqnarray*}
e_{2}(2\stackrel{4}{\longrightarrow }2) & = & \sqrt{1-\frac{1}{\sqrt{3}}}\: e(2\stackrel{2}{\longrightarrow }0)\bullet e(0\stackrel{2}{\longrightarrow }2)\\
 &  & -\frac{1}{\sqrt{6\sqrt{3}}}\: e_{1}(2\stackrel{2}{\longrightarrow }2)\bullet e_{1}(2\stackrel{2}{\longrightarrow }2)\\
 &  & -\frac{1}{3}\sqrt{\frac{3}{2}+\sqrt{3}}\: e_{1}(2\stackrel{2}{\longrightarrow }2)\bullet e_{2}(2\stackrel{2}{\longrightarrow }2)\\
 &  & -\sqrt{-\frac{4}{3}+\frac{7}{3\sqrt{3}}}\: e_{2}(2\stackrel{2}{\longrightarrow }2)\bullet e_{1}(2\stackrel{2}{\longrightarrow }2)\\
 &  & -\frac{1}{3}\sqrt{-3+2\sqrt{3}}\: e_{2}(2\stackrel{2}{\longrightarrow }2)\bullet e_{2}(2\stackrel{2}{\longrightarrow }2)\\
 &  & +\sqrt{\frac{3}{2}-\frac{5}{2\sqrt{3}}}\: e(2\stackrel{2}{\longrightarrow }4)\bullet e(4\stackrel{2}{\longrightarrow }2)
\end{eqnarray*}
We could write a similar decomposition using instead products
of paths of lengths $1$ and $3$, or $3$ and $1$, or even the trivial
ones $0$ and $4$, or $4$ and $0$.

\subsection{Unit and counit}

There is an obvious unit for the product $\bullet $, which works
in both the graded and non-graded versions of the endomorphisms of
$\mathcal{E}$. Using (\ref{eq:unit_on_EP}), and the dualization
map associated with the scalar product (see (\ref{eq:dualization_map})),
it can be written as
\begin{equation}
\mathbf{1}_{\mathcal{B}}\equiv \mathbf{1}_{\mathcal{E}}
      \otimes \sharp\left(\mathbf{1}_{\mathcal{E}}\right)
\label{eq:unit_on_End(EP)}
\end{equation}
As we already have a coproduct, we can find the counit using the axioms
it satisfies. In particular\[
\left(id\otimes \epsilon \right)\Delta (a\otimes u)=a\otimes u\]
requires\begin{equation}
\epsilon (a\otimes u)\equiv u(a)\label{eq:counit_on_End(EP)}
\end{equation}
or, equivalently, $\epsilon (\rho )=\mathrm{Tr}(\rho )$.

\subsection{Comonoidality}

The algebra $\mathcal{B} \equiv End_{\#}(\mathcal{E})$ we have defined is
not a bialgebra in the usual sense, since
\begin{eqnarray}
\Delta \mathbf{1}_{\mathcal{B}} & \neq & \mathbf{1}_{\mathcal{B}}\otimes \mathbf{1}_{\mathcal{B}}
\end{eqnarray}
therefore $\mathcal{B}$ is a {\em weak bialgebra}. It is, however, comonoidal, which
means that it satisfies both the left and right comultiplicativity
conditions of the unit \cite{Nill,BoNiSzl},\begin{eqnarray*}
\Delta ^{2}\mathbf{1}_{\mathcal{B}} & = & \left(\Delta \mathbf{1}_{\mathcal{B}}\otimes \mathbf{1}_{\mathcal{B}}\right)\bullet \left(\mathbf{1}_{\mathcal{B}}\otimes \Delta \mathbf{1}_{\mathcal{B}}\right)\\
\Delta ^{2}\mathbf{1}_{\mathcal{B}} & = & \left(\mathbf{1}_{\mathcal{B}}\otimes \Delta \mathbf{1}_{\mathcal{B}}\right)\bullet \left(\Delta \mathbf{1}_{\mathcal{B}}\otimes \mathbf{1}_{\mathcal{B}}\right)
\end{eqnarray*}
The important consequence of this property is that the category
of $End_{\#}(\mathcal{E})$-comodules is a monoidal category.

We will check explicitly the first property. Using (\ref{eq:unit_on_EP}) and
(\ref{eq:unit_on_End(EP)}) with $e_{v}^{(0)} \equiv [v]$ and $e^{(0)v}$ its dual,
the LHS becomes\begin{eqnarray*}
\Delta ^{2}\mathbf{1}_{\mathcal{B}} & = & \left(\Delta \otimes id\right)\Delta \mathbf{1}_{\mathcal{B}}\\
 & = & \sum _{v,w,x,y\in G}\left(e_{v}^{(0)}\otimes e^{(0)x}\right)\otimes \left(e_{x}^{(0)}\otimes e^{(0)y}\right)\otimes \left(e_{y}^{(0)}\otimes e^{(0)w}\right)
\end{eqnarray*}
This has to be compared with the RHS
\begin{eqnarray*}
 \left(\Delta \mathbf{1}_{\mathcal{B}}\otimes \mathbf{1}_{\mathcal{B}}\right)\bullet \left(\mathbf{1}_{\mathcal{B}}\otimes \Delta \mathbf{1}_{\mathcal{B}}\right) 
= \; \mathbf{1}_{(1)}  \otimes \left(\mathbf{1}_{(2)}\bullet \mathbf{1}_{(1)'}\right)\otimes \mathbf{1}_{(2)'} \qquad  \qquad  \qquad  \qquad  \qquad & & \\ 
\qquad = \; \sum _{\begin{array}{c}  v,w,x\\
                                                 v',w',x'\end{array}
                     }\left(e_{v}^{(0)}  \otimes  e^{(0)x}\right)
  \otimes \left[\left(e_{x}^{(0)}\otimes e^{(0)w}\right)\bullet \left(e_{v'}^{(0)}\otimes e^{(0)x'}\right)\right]  \otimes \left(e_{x'}^{(0)}\otimes e^{(0)w'}\right) & &
\end{eqnarray*}
Considering that the product in square brackets above is \begin{eqnarray*}
\left(e_{x}^{(0)}\otimes e^{(0)w}\right)\bullet \left(e_{v'}^{(0)}\otimes e^{(0)x'}\right) & = & e_{x}^{(0)}e_{v'}^{(0)}\otimes e^{(0)w}e^{(0)x'}\\
 & = & \delta _{x,v'}\, \delta _{w,x'}\, e_{x}^{(0)}\otimes e^{(0)w}
\end{eqnarray*}
we conclude that\[
\left(\Delta \mathbf{1}_{\mathcal{B}}\otimes \mathbf{1}_{\mathcal{B}}\right)\bullet \left(\mathbf{1}_{\mathcal{B}}\otimes \Delta \mathbf{1}_{\mathcal{B}}\right)=\sum _{\begin{array}{c}
 v,w,x\\
 w'\end{array}
}\left(e_{v}^{(0)}\otimes e^{(0)x}\right)\otimes \left(e_{x}^{(0)}\otimes e^{(0)w}\right)\otimes \left(e_{w}^{(0)}\otimes e^{(0)w'}\right)\]
which obviously coincides with the expression we got above for $\Delta ^{2}\mathbf{1}_{\mathcal{B}}$
after an index relabeling.

The check of the right comonoidality property is just a trivial variation
of the above. Weak multiplicativity of the counit (the {}``dual''
property) does not hold in general.

\subsection{Non-existence of an antipode}

Given an algebra or coalgebra, the unit and counit must be unique if
they exist at all, and  this is so for the weak bialgebra  
$(\mathcal{B}=End_{\#}(\mathcal{E}),\bullet )$.
One could hope to find a corresponding antipode to turn this bialgebra
into a weak Hopf algebra but this is not possible, as we will show now.

We refer the reader to \cite{Nill,BoNiSzl,BoSzl} for axioms concerning the antipode in weak Hopf algebras.
There are slight variations among
these references, for instance \cite{Nill} defines first left and
right pre-antipodes, as an intermediate step to have an antipode. This
is not relevant here, as the axioms for an antipode in any of \cite{Nill,BoNiSzl,BoSzl}
necessarily imply that $S$ must be such that
\begin{equation}
S(x_{(1)})\, x_{(2)}=\mathbf{1}_{(1)}\, \epsilon \left(x\, \mathbf{1}_{(2)}\right)\label{eq:antipode_condition}
\end{equation}
for any element $x$ of the Hopf algebra. Therefore, we can assume that
this holds for an element $\rho \in End(\mathcal{E}_{n})$
of the form\[
\rho =a\otimes u\qquad \qquad \textrm{with}\quad a=a^{(n)}\in \mathcal{E}_{n}\; ,
\quad u=u^{(n)}\in \mathcal{E}_{n}^{*}\; ,\quad n \geq 1\]
 Using $\Delta \rho =\sum _{I}\left(a\otimes e^{(n)I}\right)\otimes \left(e_{I}^{(n)}\otimes u\right)$
and replacing it in (\ref{eq:antipode_condition}) we get\[
\sum _{I}S\left(a\otimes e^{(n)I}\right)\bullet \left(e_{I}^{(n)}\otimes u\right)\]
on the LHS and\[
\sum _{v,w,x\in G}\left(e_{v}^{(0)}\otimes e^{(0)x}\right)\epsilon \left[\left(a\bullet e_{x}^{(0)}\right)\otimes \left(u\bullet e^{(0)w}\right)\right]\]
on the RHS. In this last term the sum over points $x,w$ of the graph
contributes only when $x$ is the ending point $a_{f}$ of the path
$a$, and $w$ is the ending point of (the dual of) $u$. Therefore,
we must have\begin{eqnarray*}
\sum _{I}S\left(a\otimes e^{(n)I}\right)\bullet \left(e_{I}^{(n)}\otimes u\right) & = & u(a)\, \left(\sum _{v}e_{v}^{(0)}\right)\otimes e^{(0)a_{f}}
\end{eqnarray*}
We see now that this is not possible, as the LHS gives tensor product
factors of grading $\geq n$ ---the product of whatever comes out of the
antipode times $e_{I}^{(n)}$ will always be a path of length
at least $n$, or the null element--- whereas the RHS involves paths of
length zero and is non-null in the general case. Hence, it is
not possible to find an operator $S$ which could satisfy the axiom (\ref{eq:antipode_condition}).

\subsection{The star operation}

We can define a star operation $\star$ on $Paths$ and $\mathcal{E}$
just by reversing the orientation of the paths:
\begin{eqnarray*}
p^{\star } & = & [a_{L},a_{L-1},\cdots ,a_{1},a_{0}]\, \equiv \, \tilde{p}\qquad \textrm{if}\qquad p=[a_{0},a_{1},\cdots ,a_{L}]
\end{eqnarray*}
and extending it by anti-linearity. Of course, if $e$ is essential
then $e^{\star }$ will also be essential, and a basis of $\mathcal{E}$
can always be chosen so as to have both a vector $e_{i}$ and its
conjugate in the basis, thus $e_{i}^{\star }\equiv e_{j}$ for some
$j$. 

The antilinear mapping $\star $ turns $(\mathcal{E},\bullet )$ into
a $\star $-algebra, because $P\,\star = \star \,P$; therefore\[
(a\bullet b)^{\star }=b^{\star }\bullet a^{\star }\]
and\[ 
\left(\mathbf{1}_{\mathcal{E}}\right)^{\star }=\mathbf{1}_{\mathcal{E}}\]
We can also introduce a conjugation on the algebra $\mathcal{B}=End_{\#}(\mathcal{E})$
by making use of the above one, defining\begin{eqnarray*}
\star :End_{\#}(\mathcal{E}) & \longmapsto  & End_{\#}(\mathcal{E})
\end{eqnarray*}
on monomials by
\[
\left(a\otimes u\right)^{\star }\equiv a^{\star }\otimes u^{\star }\]
This operation trivially verifies\begin{eqnarray*}
\left(\mathbf{1}_{\mathcal{B}}\right)^{\star } & = & \mathbf{1}_{\mathcal{B}}\\
\epsilon \left(\rho ^{\star }\right) & = & \overline{\epsilon \left(\rho \right)}
\end{eqnarray*}
and\[
\left(\rho \bullet \rho '\right)^{\star }=\left.\rho '\right.^{\star }\bullet \rho ^{\star }\]
To prove that\[
\Delta \left(\rho ^{\star }\right)=\left(\Delta \rho \right)^{\star \otimes \star }\]
one should only note that\[
\sum _{J}\left(e^{J}\right)^{\star }\otimes e_{J}^{\star }=\sum _{J}e^{J}\otimes e_{J}\]
for each orthonormal sub-basis $\left\{ e_{J}\right\} =\left\{ e_{J}^{(n)}\right\} $
of definite grading $n$, which holds because we can always choose
the $e_{J}$ such that $e_{J}^{\star }=e_{I}$ for some $I$. This
star operation is a normal (non-twisted) one, however it would also
be possible to introduce a twisted \cite{CoGaTr-Stars} version.

\section{Proof of the weak bialgebra compatibility condition\label{sec:weak_bialgebra_condition_proof}}

We prove in this section that, in the case of the algebra of graded endomorphisms of essential paths
$\mathcal{B}=End_{\#}(\mathcal{E})$
the condition (\ref{eq:homo_condition_graded}) holds. This condition, as we 
have seen, insures the homomorphism property of the coproduct. Some auxiliary but
relevant results are obtained first.

\subsection{Decomposition of essential paths\label{sub:decomposition-of-EP}}

An essential path of well defined endpoints $a,b$ and length $L$,\[
e=e(a\stackrel{L}{\longrightarrow }b)\]
is necessarily a linear combination \[
\sum _{p}\alpha _{p}\, p(a\stackrel{L}{\longrightarrow }b)\]
where all the $p$ are elementary paths from $a$ to $b$. Of course
we can now take $0\leq l\leq L$ and rewrite each $p$ using subpaths
of lengths $l$, $L-l$, namely $p(a\stackrel{L}{\longrightarrow }b)=p'(a\stackrel{l}{\longrightarrow }v)\, p''(v\stackrel{L-l}{\longrightarrow }b)$
for some $v\in G$, and $p',p''$ elementary too. Therefore,\[
e=\sum _{v\in G}\; \sum _{p',p''}\: \alpha _{vp'p''}\, p'(a\stackrel{l}{\longrightarrow }v)\, p''(v\stackrel{L-l}{\longrightarrow }b)\]
As $e$ is essential, in particular it must happen that $C_{k}e=0$
for $k=1,2,\cdots ,l-1$. But for these values of $k$

\[
0=C_{k}e=\sum _{v\in G}\; \sum _{p''}\; C_{k}\left(\sum _{p'}\: \alpha _{vp'p''}\, p'(a\stackrel{l}{\longrightarrow }v)\right)\, p''(v\stackrel{L-l}{\longrightarrow }b)\]
and using the linear independence of the elementary paths $p''$ we
see that for each of the possible $p''$ the linear combination in
parentheses must be essential:\[
\sum _{p'}\: \alpha _{vp'p''}\, p'(a\stackrel{l}{\longrightarrow }v)\equiv \sum _{i}\beta _{vip''}\, e_{i}^{\prime }(a\stackrel{l}{\longrightarrow }v)\]
Here the index $i$ runs over a basis of essential paths of definite
endpoints $a,v$ and length $l$. Getting this back into $e$, we
get\[
e=\sum _{v\in G}\; \sum _{i,p''}\: \beta _{vip''}\, e_{i}^{\prime }(a\stackrel{l}{\longrightarrow }v)\, p''(v\stackrel{L-l}{\longrightarrow }b)\]
We now use that $C_{k}e=0$ for $k=l+1,\cdots ,L-1$, so\[
0=C_{k}e=\sum _{v,i}\, e_{i}^{\prime }(a\stackrel{l}{\longrightarrow }v)\: C_{k-l}\left(\sum _{p''}\: \beta _{vip''}\, p''(v\stackrel{L-l}{\longrightarrow }b)\right)\]
and due to the linear independence of the basis $\left\{ e_{i}^{,}\right\} $
of essential paths we conclude again that for any value of $i$ and $v$ the
term in parentheses must be essential:\[
\sum _{p''}\: \beta _{vip''}\, p''(v\stackrel{L-l}{\longrightarrow }b)\equiv \sum _{j}\gamma _{vij}\, e_{j}^{\prime \prime }(v\stackrel{L-l}{\longrightarrow }b)\]
Putting this back into $e$, and using $P(e)=e$, we obtain the desired factorization:
\begin{eqnarray*}
e &=& \sum _{v,i,j}\: \gamma _{vij}\, P \left( 
               e_{i}^{\prime }(a\stackrel{l}{\longrightarrow }v) \, 
               e_{j}^{\prime \prime }(v\stackrel{L-l}{\longrightarrow }b) \right) \\
  &=& \sum _{v,i,j}\: \gamma _{vij}\, 
               e_{i}^{\prime }(a\stackrel{l}{\longrightarrow }v) \bullet
               e_{j}^{\prime \prime }(v\stackrel{L-l}{\longrightarrow }b)
\end{eqnarray*}
The cases $l=0,L$ are completely trivial. We can formulate this intermediate
result as a lemma.\\

\paragraph*{Lemma}

Any essential path $e(a\stackrel{L}{\longrightarrow }b)$ of well
defined endpoints $a,b$ and length $L$ can be decomposed, for any
fixed given positive value $l<L$, as a linear combination of products
of shorter essential paths
\begin{equation}
e(a\stackrel{L}{\longrightarrow }b)=\sum _{v,i,j}\: \gamma _{vij}\, 
               e_{i}^{\prime }(a\stackrel{l}{\longrightarrow }v) \bullet
               e_{j}^{\prime \prime }(v\stackrel{L-l}{\longrightarrow }b)
\label{EP-decomposition}
\end{equation}
where the sum extends over all possible points $v$ of the graph which
can be reached from $a$ and $b$ with essential paths of length $l$
and $L-l$, respectively. If we assume that both (sub)basis $\left\{ e_{i}^{\prime }\right\} $
and $\left\{ e_{j}^{\prime \prime }\right\} $ are orthonormal then also
\begin{eqnarray}
\sum _{v,i,j}\: \left|\gamma _{vij}\right|^{2} & = & \left\Vert e\right\Vert ^{2}\label{EP-norm}
\end{eqnarray}
$\blacksquare$

Note that the decomposition (\ref{EP-decomposition}) can be used to build the
essential paths recursively. With regard to the dimensionality of this space,
remark that when $G$ is a Dynkin diagram of type ADE, the following result
(that we do not prove here) is known:
The vector space spanned by the vertices $a,b,\cdots$ of $G$ is a module
over the graph algebra of $A_{n}$, where $n+1$ is the Coxeter number of $G$
and $A_{n}$ is the commutative algebra with generators
$N_{0}, N_{1}, \cdots, N_{n-1}$ obeying the following relations:
$N_{0}$ is the unit, $N_{1}$ is the (algebraic) generator with
$N_{1} N_{p} = N_{p-1} + N_{p+1}$, if $p < n-1$, and
$N_{1} N_{n-1} = N_{n-2}$. If $s$ denotes the number of vertices
of $G$, this module action is encoded by $n$ matrices $F_{p}$ of size
$s \times s$. They are related to the previous generators by
$N_{p} a = \sum_{b} (F_{p})_{ab} b$. The number of essential paths
of length $p$ on the graph $G$ is equal to the
sum of the matrix elements of $F_{p}$.

\subsection{The weak bialgebra condition}

The coefficients $m_{nI,mJ}^{(n+m)K}$ that enter the weak bialgebra
condition (\ref{eq:graded_sum_mm_condition}) are just the components
of products $e_{I}^{(n)}\bullet e_{J}^{(m)}$ of essential paths of
lengths $n,m$ respectively, along the directions $e_{K}^{(n+m)}$.
Using a more explicit notation than above, the non-trivial contributions
are\begin{eqnarray*}
m_{e_{n}(a\stackrel{l}{\longrightarrow }c)\, ,\, e_{r}(c\stackrel{L-l}{\longrightarrow }b)}^{e_{k}(a\stackrel{L}{\longrightarrow }b)} & \equiv  & \left\langle e_{k}\, ,\, e_{n}\bullet e_{r}\right\rangle
 \: = \: \left\langle e_{k}\, ,\, e_{n}\, e_{r}\right\rangle
\end{eqnarray*}
where we have used the definition (\ref{eq:product-on-EP}) for the
product, self-adjointness of the operator $P$, and the fact that
$e_{k}$ is essential so $P(e_{k})=e_{k}$. Taking $e=e_{k}$ in the
decomposition (\ref{EP-decomposition}) we can now write\begin{eqnarray*}
m_{e_{n}\, ,\, e_{r}}^{e_{k}} & = & \sum _{v,i,j}\: \gamma _{vij}^{(k)}\, \left\langle e_{i}^{\prime }(a\stackrel{l}{\longrightarrow }v)\, ,\, e_{n}(a\stackrel{l}{\longrightarrow }c)\right\rangle \, \left\langle e_{j}^{\prime \prime }(v\stackrel{L-l}{\longrightarrow }b)\, ,\, e_{r}(c\stackrel{L-l}{\longrightarrow }b)\right\rangle \\
 & = & \sum _{v,i,j}\: \gamma _{vij}^{(k)}\, \delta _{vc}\, \delta _{in}\, \delta _{jr}
 \: = \: \gamma _{cnr}^{(k)}
\end{eqnarray*}
Therefore the coefficients $\gamma _{cnr}^{(k)}$ that enter the decomposition
of $e_{k}$ are the same that those involved in the product. The weak
bialgebra condition (\ref{eq:graded_sum_mm_condition}) reduces now
to the orthonormality condition (\ref{EP-norm}) of the $e_{k}$ , that is
\begin{eqnarray}
\sum _{nr}\overline{m_{e_{n}(a\stackrel{l}{\longrightarrow }c)\, ,\, e_{r}(c\stackrel{L-l}{\longrightarrow }b)}^{e_{k}(a\stackrel{L}{\longrightarrow }b)}}\,
m_{e_{n}(a\stackrel{l}{\longrightarrow }c)\, ,\, e_{r}(c\stackrel{L-l}{\longrightarrow }b)}^{e_{k'}(a\stackrel{L}{\longrightarrow }b)}
& = & \sum _{c,n,r}\: \overline{\gamma _{cnr}^{(k)}}\, \gamma _{cnr}^{(k')} \nonumber\\
& = & \left\langle e_{k}\, ,\, e_{k'}\right\rangle 
\: = \: \delta _{kk'} \nonumber
\end{eqnarray}

\section{Comparison of the two bialgebra structures for the $A_{2}$ diagram}

The graph $A_{2}$ gives rise to the simplest non-trivial example, an
8-dimensional algebra  (whereas $A_{3}$ already produces a 34-dimensional
one). It consists of two points and one (bi-oriented) edge. The only essential
paths are: $a_{1}\equiv [1]$, $a_{2}\equiv [2]$, and the right and
left oriented paths $r\equiv [1,2]$ and $l\equiv [2,1]$ respectively.

We shall compare, for this example, the two bialgebra structures mentioned in the text.
The first, the graded one, is a weak bialgebra, semi-simple but not co-semi-simple. 
The second, the filtrated one, is a weak Hopf algebra; it is both simple and co-semi-simple.

\subsection{The graded bialgebra structure}

The products in $\mathcal{E}(A_{2})$ (corresponding to (\ref{eq:m_A_basis})) are:
\begin{eqnarray*}
a_{i}\bullet a_{j}=\delta _{ij}a_{i} & \qquad  & r^{2}=l^{2}=r\bullet l=l\bullet r=0\\
a_{1}\bullet r=r\bullet a_{2}=r &  & a_{2}\bullet r=r\bullet a_{1}=0\\
a_{2}\bullet l=l\bullet a_{1}=l &  & a_{1}\bullet l=l\bullet a_{2}=0
\end{eqnarray*}

\noindent The dual operation in $\mathcal{E}(A_{2})$, the coproduct 
corresponding to (\ref{eq:D_A_basis})), is 
\begin{eqnarray*}
Da_{1} = a_{1}\otimes a_{1} & \qquad & 
         Da_{2} = a_{2}\otimes a_{2} \\
Dr     = a_{1}\otimes r+r\otimes a_{2} & \qquad  &
         Dl     = a_{2}\otimes l+l\otimes a_{1}
\end{eqnarray*}

\noindent
Now we consider $E\equiv End_{\#}(\mathcal{E}(A_{2}))$: we call $\rho _{ij}$
the endomorphism of paths of length zero taking $a_{j}$ into $a_{i}$, 
which we also identify using the map $\sharp$ as 
$\rho _{ij}=a_{i}\otimes a_{j}$. We also have the 
$\rho _{rr},\rho _{rl},\rho _{lr},\rho _{ll}$
acting on the space of paths of length $1$. Thus $E$ has dimension $8$
as a vector space.

The product in $E$ is the usual composition product, so 
\begin{eqnarray*}
\rho _{ij}\circ \rho _{kl} & = & \delta _{jk}\rho _{il}\qquad \qquad i,j=1,2\\
\rho _{ij}\circ \rho _{**}=\rho _{**}\circ \rho _{ij} & = & 0\qquad \qquad \qquad *=r,l\\
\rho _{d_{1}d_{2}}\circ \rho _{d_{3}d_{4}} & = & \delta _{d_{2}d_{3}}\rho _{d_{1}d_{4}}\qquad d_{i}=r,l
\end{eqnarray*}
Obviously, $(E, \circ)$ is the direct sum of two subalgebras, namely
$End(\mathcal{E}_{0,1}(A_{2}))$, the endomorphisms of paths of length $i$, 
both isomorphic to $M_{2x2}(\mathbb{C})$.

Regarding the coproduct on $E$, remember that for the graded case we defined\[
\Delta \rho =(P\otimes P)(1\otimes \tau \otimes 1)(D_{A}\otimes D_{A^{*}})\rho \]
In our present example this implies
\begin{eqnarray*}
\Delta \rho _{ij} & = & \rho _{ij}\otimes \rho _{ij}\\
 &  & \\
\Delta \rho _{rr} & = & \rho _{11}\otimes \rho _{rr}+\rho _{rr}\otimes \rho _{22}\\
\Delta \rho _{ll} & = & \rho _{22}\otimes \rho _{ll}+\rho _{ll}\otimes \rho _{11}\\
\Delta \rho _{rl} & = & \rho _{12}\otimes \rho _{rl}+\rho _{rl}\otimes \rho _{21}\\
\Delta \rho _{lr} & = & \rho _{21}\otimes \rho _{lr}+\rho _{lr}\otimes \rho _{12}
\end{eqnarray*}
Indeed, in the first case, for example, the calculation reads
\begin{eqnarray*}
\Delta \rho _{rr} & = & \Delta (r\otimes r)=(P\otimes P)(1\otimes \tau \otimes 1)\left((a_{1}\otimes r+r\otimes a_{2})\otimes (a_{1}\otimes r+r\otimes a_{2})\right)\\
 & = & (P\otimes P)\left(a_{1}\otimes a_{1}\otimes r\otimes r+r\otimes r\otimes a_{2}\otimes a_{2}+a_{1}\otimes r\otimes r\otimes a_{2}+r\otimes a_{1}\otimes a_{2}\otimes r\right)\\
 & = & a_{1}\otimes a_{1}\otimes r\otimes r+r\otimes r\otimes a_{2}\otimes a_{2}=\rho _{11}\otimes \rho _{rr}+\rho _{rr}\otimes \rho _{22}
\end{eqnarray*}
because the terms $a_{1}\otimes r\otimes r\otimes a_{2}+r\otimes a_{1}\otimes a_{2}\otimes r$
do not belong to $E\otimes E$ and get projected out by the operator
$P\otimes P$. It is easy to check that $\Delta $ is both coassociative
and an algebra homomorphism for the product $\circ $. Therefore,
$E$ is a bialgebra. The element $\one=\rho _{11}+\rho _{22}+\rho _{rr}+\rho _{ll}$
is a unit for $\circ $ but its coproduct is not $\one \otimes \one$.

If we declare the elementary paths $a_1, a_2, r, l$ orthonormal, we obtain 
an induced scalar product on the space of endomorphisms. We can use it to 
map the above coproduct to a product that we call $\bullet$. 
 
The first algebra (product $\circ$) is  isomorphic, by construction, with the semi-simple algebra $M_2(\mathbb{C}) \oplus M_2(\mathbb{C})$. 
The matrix units,  or "elementary matrices", are realized as follows. Each entry denotes a single matrix unit (replace the chosen generator by $1$ and set the others entries to zero):
{\scriptsize
$
\begin{array}{ccc}
\left(
    \begin{array}{cc}
     \rho_{11} &  \rho_{12} \\
        {}&{} \\
      \rho_{21} &  \rho_{22}
    \end{array}
\right)
\oplus
\left(
     \begin{array}{cc}
    \rho_{ rr} & \rho_{ rl} \\
        {}&{} \\
    \rho_{ lr} &  \rho_{ll}
    \end{array}
\right)
\end{array}
$
}

The graded algebra (product $\bullet$) is not semi-simple. It can be realized\footnote{So, there are four projective irreducible modules and the radical is $\mathbb{C} \oplus \mathbb{C} \oplus \mathbb{C} \oplus \mathbb{C}$ } as a direct sum of two algebras of matrices $2\times 2$ with entries in the ring of Grassman numbers  with generators $\{1, \theta\}$, $\theta^2=0$. Indeed, the basis vectors $\{ \rho_{11}, \rho_{rr}, \rho_{ll}, \rho_{22} \}$ generate an algebra isomorphic with {\scriptsize $ \left(
    \begin{array}{cc}
     a & b\, \theta \\
        {}&{} \\
    c \, \theta & d
    \end{array}
\right)
$},  where $a,b,c,d$ are complex numbers.
Vectors $\{  \rho_{12}, \rho_{rl}, \rho_{lr}, \rho_{21} \}$ generate another copy of the same four-dimensional algebra.  The eight generators can be realized as (dots stand for the number $0$): 

{\scriptsize
$
\begin{array}{cc}
\rho_{11} = \begin{array}{ccc}
\left(
    \begin{array}{cc}
     1& .  \\
        . & . \\
    \end{array}
\right) 
\oplus 
\left(
    \begin{array}{cc}
     . & .  \\
        . & . \\
    \end{array}
\right) 
 \end{array}
 &
 \rho_{rr} = \begin{array}{ccc}
\left(
    \begin{array}{cc}
     . & \theta  \\
        . & . \\
    \end{array}
\right) 
\oplus 
\left(
    \begin{array}{cc}
     . & .  \\
        . & . \\
    \end{array}
\right) 
 \end{array} 
 \\
 \rho_{ll} = \begin{array}{ccc}
\left(
    \begin{array}{cc}
     . & .  \\
        \theta & . \\
    \end{array}
\right) 
\oplus 
\left(
    \begin{array}{cc}
     . & .  \\
        . & . \\
    \end{array}
\right) 
 \end{array}
 &
 \rho_{22} = \begin{array}{ccc}
\left(
    \begin{array}{cc}
     . & .  \\
        . & 1 \\
    \end{array}
\right) 
\oplus 
\left(
    \begin{array}{cc}
     . & .  \\
        . & . \\
    \end{array}
\right) 
 \end{array}
 \\
 \rho_{12} = \begin{array}{ccc}
\left(
    \begin{array}{cc}
     . & .  \\
        . & . \\
    \end{array}
\right) 
\oplus 
\left(
    \begin{array}{cc}
     . & -\theta \\
         \theta & 1  \\
    \end{array}
\right) 
 \end{array}
 &
 \rho_{rl} = \begin{array}{ccc}
\left(
    \begin{array}{cc}
     . & .   \\
        . & . \\
    \end{array}
\right) 
\oplus 
\left(
    \begin{array}{cc}
     . & .  \\
       \theta & . \\
    \end{array}
\right) 
 \end{array} 
 \\
 \rho_{lr} = \begin{array}{ccc}
\left(
    \begin{array}{cc}
     . & .  \\
        . & . \\
    \end{array}
\right) 
\oplus 
\left(
    \begin{array}{cc}
     . &  \theta \\
        . & . \\
    \end{array}
\right) 
 \end{array}
 &
 \rho_{21} = \begin{array}{ccc}
\left(
    \begin{array}{cc}
     . & .  \\
        . & . \\
    \end{array}
\right) 
\oplus 
\left(
    \begin{array}{cc}
     1 & \theta \\
       - \theta & . \\
    \end{array}
\right) 
 \end{array} 
  \end{array}
$
}

\subsection{The filtrated bialgebra structure}

The filtrated bialgebra structure associated with $A_2$ (see also \cite{Coque-Cocoyoc}) uses the same composition product $\circ$ but
the second product $\star$ is different from $\bullet$. 
Actually, the case $A_2$ is rather special, in the following sense:
there exists an associative structure (call it also $\star $) on the space of essential paths $\mathcal{E}(A_{2})$ such that  the filtrated algebra structure that we consider on the eight dimensional space $E$ coincides with the tensor square of the later. This is (unfortunately) not so for other ADE diagrams, not even for the $A_N$ when $N>2$.
The product $\star$ on $\mathcal{E}(A_{2})$  is :

\begin{eqnarray*}
a_{i}\star a_{j}=\delta _{ij}a_{i} & \qquad  & r^{2}=l^{2}= 0 \\
{} & \qquad & r\star l= a_1 \, ,\quad  l\star r= a_2\\
a_{1}\star r=r\star a_{2}=r &  & a_{2}\star r=r\star a_{1}=0\\
a_{2}\star l=l\star a_{1}=l &  & a_{1}\star l=l\star a_{2}=0
\end{eqnarray*}
Comparing with the multiplication $\bullet$ of the previous section, we see that the difference lies in the values of  $r\star l$ and $l\star r$ that, here, do not vanish.
The product $\star$ in $E$ is:
$$ (u \otimes v) \star (u'  \otimes  v')  \doteq (u \star u')  \otimes  (v \star v') $$

It is easy to write the multiplication table and to see that this algebra is semi-simple and isomorphic, like $(E, \circ)$, with the direct sum of two full matrix algebras $2\times 2$ over the complex numbers. However, the eight generators are represented in a very different way. With the same reading convention as before, the
matrix units are given  by:
{\scriptsize
$
\begin{array}{ccc}
\left(
    \begin{array}{cc}
    \rho_{11}  &  \rho_{rr }\\
        {}&{} \\
      \rho_{ll } &  \rho_{22}
    \end{array}
\right)
\oplus
\left(
     \begin{array}{cc}
     \rho_{12} & \rho_{ rl} \\
        {}&{} \\
      \rho_{lr} &  \rho_{21}
    \end{array}
\right)
\end{array}
.$
}
The corresponding coproducts (compare with the previous section) read as follow:  $\Delta \rho_{u,v}$, when 
$u,v = r,l$ are as before, but the $\Delta \rho_{i,j}$, $i,j = 1,2$ are different\footnote{Notice that
$\Delta \one = (11+ll)\otimes (11 + rr) + (rr+22) \otimes (ll + 22)$}:
\begin{eqnarray*}
\Delta \rho_{11} &=& \rho_{11}  \otimes  \rho_{11}  + rr \otimes ll \\
\Delta \rho_{12} &=&\rho_{12} \otimes  \rho_{12} + rl \otimes lr \\
\Delta \rho_{21} &=& \rho_{21} \otimes  \rho_{21} + lr \otimes rl \\
\Delta \rho_{22} &=&\rho_{22} \otimes \rho_{22} + ll \otimes rr
\end{eqnarray*}

\appendix

\section{The bialgebra of endomorphisms of an algebra}

We describe the (weak) bialgebra structure of the space of endomorphisms
of the algebra $\mathcal{E}$. Actually, only the fact that $\mathcal{E}$ 
possesses an algebra structure is needed here, so we may start from an
arbitrary algebra that we call $A$.

\subsection{$A$ and $A^{*}$\label{sub:A_and_A*}}

Take $A$ an associative algebra, with or without unit, and finite
dimensional. Call $m_{A}:A\otimes A\rightarrow A$ its product. Now
introduce the linear dual vector space $A^{*}=\{A\stackrel{lin}{\longrightarrow }\mathbb{C}\}$,
which can be automatically endowed with a (coassociative) coalgebra
structure in the standard way. We call $D_{A^{*}}:A^{*}\rightarrow A^{*}\otimes A^{*}$
its coproduct (of course, $D_{A^{*}}(u)(a\otimes b)=u(ab)$). Now
\emph{choose} a scalar product $\left\langle \: ,\: \right\rangle $
on $A$: this defines an antilinear isomorphism%
\footnote{$\sharp$ goes from $A$ to $A^{*}$, but in the same way one can define
$\sharp^{*}$ from $A^{*}$ to $A^{**}$. As both $A$ and $A^{**}$ can
be identified in the finite dimensional case, $\sharp$ is invertible.%
} $\sharp$ between $A$ and $A^{*}$, given by
\begin{eqnarray}
\sharp \, : \, A & \longrightarrow  & A^{*}\label{eq:dualization_map}\\
a & \longrightarrow  & \sharp(a)=\left\langle a,\: \right\rangle \nonumber
\end{eqnarray}
Its inverse is usually called $\flat \equiv \sharp^{-1}$. Use this
identification $\sharp$ to define a comultiplication on $A$
\begin{eqnarray}
D_{A} & : & A\longmapsto A\otimes A\label{eq:coproduct_A}\\
D_{A}(a) & \equiv  & \left(\flat \otimes \flat \right)D_{A^{*}}(\sharp(a))\nonumber
\end{eqnarray}
and its dual, a product $m_{A^{*}}$ on $A^{*}$:\begin{eqnarray}
m_{A^{*}} & : & A^{*}\otimes A^{*}\longmapsto A^{*}\label{eq:product_A*}\\
m_{A^{*}}(u\otimes v) & \equiv & \sharp\circ m_{A} \circ (\flat\otimes\flat) (u \otimes v)
= \sharp \left(\flat(u)\flat(v)\right)\nonumber
\end{eqnarray}
That is, we choose $m_{A^{*}}$ in such a way that $\sharp$ becomes an algebra
homomorphism. Note that if $\left\{ e_{i}\right\} $ is an orthonormal basis of
$A$ and \begin{eqnarray}
e_{i}e_{j} & = & \sum _{k}m_{ij}^{k}\: e_{k}\label{eq:m_A_basis}
\end{eqnarray}
then using $e^{i}\equiv \sharp(e_{i})$ we have
\begin{eqnarray}
D_{A^{*}}(e^{k}) & = & \sum _{ij}m_{ij}^{k}\: e^{i}\otimes e^{j}\label{eq:D_A*_basis}\\
e^{i}e^{j}\equiv m_{A^{*}}(e^{i}\otimes e^{j}) & = & \sum _{k}\overline{m_{ij}^{k}}\: e^{k}\label{eq:m_A*_basis}\\
D_{A}(e_{k}) & = & \sum _{ij}\overline{m_{ij}^{k}}\: e_{i}\otimes e_{j}\label{eq:D_A_basis}
\end{eqnarray}
Having the same coefficients $m_{ij}^{k}$ as $m_{A}$, the operations
$D_{A}$ and $m_{A^{*}}$ are automatically (co)associative.
Neither $A$ nor $A^{*}$ are a priori bialgebras, so there
is no reason for $D_{A}$ or $D_{A^{*}}$ to be algebra homomorphisms
with respect to $m_{A}$ or $m_{A^{*}}$.

\subsection{$End(A)$ - the non-graded case\label{sub:End(A)}}

Let us now see what we can do on $End(A)$. Notice that we are not
considering any graduation on $A$ (in this subsection, elements of 
$End(A)$ are general endomorphisms). We know that\[
End(A)\simeq A\otimes A^{*}\:\];
the algebra structure of $End(A)$ is given by the associative
composition product
\begin{equation}
\rho \circ \rho '=(a\otimes u)\circ (a'\otimes u')\equiv a\otimes u(a')\: u'  \qquad 
\textrm{whenever}\; \rho =a\otimes u\, ,\; \rho '=a'\otimes u'\label{eq:composition_product}
\end{equation}
and uses nothing more than the vector space structure of $A$.

If $\Phi $, $\Phi '$ are vectorial
homomorphisms from a coalgebra $A$ to an algebra $B$, one can define
a convolution product; whereas if $\Phi $ is a vectorial homomorphism
from an algebra $A$ to a coalgebra $B$, a convolution coproduct
can be given on $\Phi $. This comes from the fact that $Hom (A,B)\simeq B\otimes A^{*}$.
Coming back to our case, taking $B=A$ equipped with both a
product $m_{A}$ and a coproduct%
\footnote{either because we started from an algebra $A$, selected a scalar product,
and applied the procedure of the previous subsection, or because we
started from a genuine bialgebra $A$.%
} $D_{A}$ (ergo also their dual operations $D_{A^{*}}$, $m_{A^{*}}$
on $A^{*}$), the convolution product\footnote{Warning: calling $\bullet$ 
a convolution product may be misleading since the already mentioned filtrated
multiplication is also of the same type. The reader will certainly understand
which is which from the context.} is\begin{eqnarray}
(\rho \bullet \rho ')(a) & \equiv  & \rho (a_{1}).\rho '(a_{2})\qquad \qquad \textrm{where}\; D_{A}(a)=a_{1}\otimes a_{2}\label{eq:convolution_product}
\end{eqnarray}
This is just the natural multiplication in the tensor product of algebras of $A$ and $A^{*}$:
\begin{eqnarray}
(a\otimes u)\bullet (a'\otimes u') & = & aa'\otimes uu'\label{eq:convolution_product_explicit}\\
\rho \bullet \rho ' & = & \left(m_{A}\otimes m_{A^{*}}\right)(1\otimes \tau \otimes 1)(\rho \otimes \rho ')\nonumber
\end{eqnarray}
where $a, a' \in A$, $u,u' \in A^{*}$ and $\tau $ is the twist permuting
two factors of a tensor product.
In particular if we start from a vector space $A$ endowed with both an algebra
and a coalgebra structure (it may be, or not, a bialgebra), the above construction
gives two distinct multiplicative structures to the space $End(A)$ ---the composition
product $\circ$ and the product  $\bullet$. Equivalently,
it gives two distinct comultiplicative structures to the space $End(A^{*})$. 
Now, if  we want to consider $End(A)$ both
as an algebra and a coalgebra, one has somehow to identify $End(A^{*})$ with
$End(A)$, and, for this reason, we need to choose some scalar product.

Let us therefore consider an algebra $A$ endowed with some given
scalar product, and ``dualize'' one of the two products on $End(A)$ ---either $\circ $ or
$\bullet $--- to get a coproduct on $End(A)^{*}$. Finally, we map the latter to a
comultiplication on $End(A)$ simply by using the isomorphism $\flat\otimes \sharp$.
For later convenience we choose to dualize the composition product
$\circ $. In this way we obtain the {}``composition'' coproduct
$\Delta $ on $End(A)$,\begin{eqnarray}
\Delta  & : & End(A)\longmapsto End(A)\otimes End(A)\label{eq:composition_coproduct}\\
 &  & \nonumber \\
\Delta (a\otimes u) & = & \sum _{i}\left(a\otimes e^{i}\right)\otimes \left(e_{i}\otimes u\right)\quad \in \left(A\otimes A^{*}\right)\otimes \left(A\otimes A^{*}\right)\nonumber
\end{eqnarray}
where the sum runs over an orthonormal basis of $A$ (for the chosen scalar product) as in the previous
subsection. This coproduct $\Delta $ is trivially coassociative,
and the product $\bullet $ is associative due to the corresponding
properties of $m_{A}$ and $m_{A^{*}}$.

Now we want $End(A)$ to be a bialgebra but the comultiplication $\Delta $ that we just considered has a priori no reason to
be an algebra homomorphism%
\footnote{This would still be the case even if $A$ were a true bialgebra. However,
we are not making this hypothesis here.%
} for $\bullet $, ie, in general $\Delta (\rho \bullet \rho ')\neq \Delta \rho \, (\bullet \otimes \bullet )\, \Delta \rho '$.
Let us, however, analyze the terms separately:
\begin{eqnarray}
\Delta (\rho \bullet \rho ') & = & \Delta ((a\otimes u)\bullet (a'\otimes u'))=\Delta (aa'\otimes uu')\label{eq:Delta_of_product}\\
 & = & \sum _{i}\left(aa'\otimes e^{i}\right)\otimes \left(e_{i}\otimes uu'\right)\nonumber 
\end{eqnarray}
On the other hand,\begin{eqnarray*}
\Delta \rho \, (\bullet \otimes \bullet )\, \Delta \rho ' & = & \sum _{ij}\left[\left(a\otimes e^{i}\right)\otimes \left(e_{i}\otimes u\right)\right]\, (\bullet \otimes \bullet )\, \left[\left(a'\otimes e^{j}\right)\otimes \left(e_{j}\otimes u'\right)\right]\\
 & = & \sum _{ij}\left(aa'\otimes e^{i}e^{j}\right)\otimes \left(e_{i}e_{j}\otimes uu'\right)
\end{eqnarray*}
Using the explicit expressions (\ref{eq:m_A_basis}),(\ref{eq:m_A*_basis})
this becomes
\begin{eqnarray}
\Delta \rho \, (\bullet \otimes \bullet )\, \Delta \rho ' & = & \sum _{kl}\left(\sum _{ij}\overline{m_{ij}^{k}}\, m_{ij}^{l}\right)\left(aa'\otimes e^{k}\right)\otimes \left(e_{l}\otimes uu'\right)\label{eq:product_of_Deltas}
\end{eqnarray}
$(End(A),\bullet ,\Delta )$ would be a bialgebra only if (\ref{eq:Delta_of_product})
and (\ref{eq:product_of_Deltas}) coincide. As the elements $\rho $
and $\rho '$ (in fact $a,a',u,u'$) can be chosen arbitrarily, this
requires\[
\sum _{kl}\left(\sum _{ij}\overline{m_{ij}^{k}}\, m_{ij}^{l}\right)e^{k}\otimes e_{l}=\sum _{k}e^{k}\otimes e_{k}\]
namely\begin{equation}
\sum _{ij}\overline{m_{ij}^{k}}\, m_{ij}^{l}=\delta ^{kl}\qquad \qquad \forall k,l\label{eq:sum_mm_condition}\end{equation}
This requirement can be rewritten also as \begin{equation}
m_{A}\left(D_{A}(a)\right)=a\qquad \qquad \forall a\in A\label{eq:m_D_condition}\end{equation}
It is a necessary condition for $(End(A),\bullet ,\Delta )$
to be a bialgebra. One may be surprised to see that this condition does not
seem to involve the chosen scalar product on $A$... but it does, since $D_A$ itself involves it (and if there
would be no chosen scalar product, the composition coproduct $\Delta$  would only be defined
on the dual of $End(A)$, so that one could not even ask for this compatibility requirement).
Remark: we used here the composition coproduct $\Delta $ and
the convolution product $\bullet $, but exactly the same can be done
in the dual picture, ie, taking the convolution coproduct $\Delta _{\bullet }$
and the composition product $\circ $. The resulting condition is
exactly the same.

\subsection{$End_{\#}(A)$ - the graded case\label{sub:graded-End(A)}}

In this subsection we particularize the above discussion to the case of a 
\emph{graded algebra} $A$, where $A=\bigoplus _{n}A_{n}$ for the underlying 
vector space, and $m_{A}:A_{n}\otimes A_{m}\rightarrow A_{n+m}$.
Now we restrict the endomorphisms to be grade-preserving:
\begin{eqnarray*}
End_{\#}(A) & = & \bigoplus _{n}End(A_{n}) \\
& \stackrel{iso}{\simeq} & \bigoplus_{n}A_{n}\otimes\left( A_{n}\right)^{*}
\end{eqnarray*}
Hence $End_{\#}(A)$ is a graded space, and the composition product
preserves this grading, 
\[
\circ \, :\, End(A_{n})\otimes End(A_{k})\longmapsto \delta _{nk} \, End(A_{n})
\]

\noindent
As $A$ is graded, its dual $A^{*}$ can be decomposed as
$A^{*}=\bigoplus _{n}\left(A_{n}\right)^{*}$. We can also write, for instance,
\[
D_{A}\, :\, A_{n}\longmapsto \bigoplus _{k=0,\cdots ,n}A_{n-k}\otimes A_{k}
\]

\noindent
The convolution product (\ref{eq:convolution_product})
turns $End_{\#}(A)$ into a graded algebra, \[
\bullet \, :\, End(A_{n})\otimes End(A_{k})\longmapsto End(A_{n+k})\]
as it is easy to see from the explicit expression (\ref{eq:convolution_product_explicit}):
take $a,\, u$ of grading $n$, and $a',\, u'$ of grading $k$, thus
both $aa'$ and $uu'$ have grading $n+k$.

The composition coproduct defined by (\ref{eq:composition_coproduct}) has to
be restricted with a projector \begin{eqnarray*}
P_{\#} & : & End(A)\longmapsto End_{\#}(A)\\
 &  & \\
P_{\#}\left(a^{(n)}\otimes u^{(k)}\right) & = & \delta _{nk}\, a^{(n)}\otimes u^{(k)}\qquad \qquad \forall n,k,\; a^{(n)}\in A_{n}\, ,\; u^{(k)}\in A_{k}^{*}
\end{eqnarray*}
if we want its image to be inside $End_{\#}(A)\otimes End_{\#}(A)$. This
comes from the fact that the dual product may be defined just on elements
of $\bigoplus _{n}A_{n}^{*}\otimes A_{n}$ or extended to the whole
$A^{*}\otimes A$. Therefore\begin{eqnarray}
\Delta  & : & End_{\#}(A)\longmapsto End_{\#}(A)\otimes End_{\#}(A)\label{eq:coproduct_on_E}\\
\Delta (a\otimes u) & = & \sum _{i}P_{\#}\left(a\otimes e^{i}\right)\otimes P_{\#}\left(e_{i}\otimes u\right)\nonumber 
\end{eqnarray}
where we assumed the basis elements $e_{i}$ to have definite grade.
Writing explicitly the grading of each $e_{i}$ as $e_{I}^{(n)}\in A_{n}$
(now $I$ runs over a basis of $A_{n}$) and using the projectors
$P_{\#}$ we get

\begin{equation}
\Delta \left(a^{(n)}\otimes u^{(n)}\right)=\sum _{I}\left(a^{(n)}\otimes e^{(n)I}\right)\otimes \left(e_{I}^{(n)}\otimes u^{(n)}\right)\qquad \qquad a^{(n)}\in A_{n}\; ,\quad u^{(n)}\in A_{n}^{*}\label{eq:graded_composition_coproduct}\end{equation}

As before, $\Delta $ and $\bullet $ are co/associative, but not
necessarily compatible (we want $\Delta $ to be an algebra
homomorphism for $\bullet $). The necessary
condition is  a slight modification of the one presented for the
non-graded case in the previous section. Take two endomorphisms of
definite grade, $\rho _{n}=a^{(n)}\otimes u^{(n)}\in A_{n}\otimes A_{n}^{*}$
and $\rho _{k}^{\prime }=a^{\prime (k)}\otimes u^{\prime (k)}\in A_{k}\otimes A_{k}^{*}$,
and redo (\ref{eq:Delta_of_product}) and (\ref{eq:product_of_Deltas})
explicitly incorporating the grading in the notation. Then\begin{eqnarray*}
\Delta (\rho \bullet \rho ') & = & \Delta \left(\left(a^{(n)}\otimes u^{(n)}\right)\bullet \left(a^{\prime (k)}\otimes u^{\prime (k)}\right)\right)=\Delta \left(a^{(n)}a^{\prime (k)}\otimes u^{(n)}u^{\prime (k)}\right)\\
 & = & \sum _{I}\left(a^{(n)}a^{\prime (k)}\otimes e^{(n+k)I}\right)\otimes \left(e_{I}^{(n+k)}\otimes u^{(n)}u^{\prime (k)}\right)\\
 &  & \\
\Delta \rho \, (\bullet \otimes \bullet )\, \Delta \rho ' & = & \sum _{IJ}\left[\left(a^{(n)}\otimes e^{(n)I}\right)\otimes \left(e_{I}^{(n)}\otimes u^{(n)}\right)\right]\, (\bullet \otimes \bullet )\\
 &  & \qquad \qquad \left[\left(a^{\prime (k)}\otimes e^{(k)J}\right)\otimes \left(e_{J}^{(k)}\otimes u^{\prime (k)}\right)\right]\\
 & = & \sum _{IJ}\left(a^{(n)}a^{\prime (k)}\otimes e^{(n)I}e^{(k)J}\right)\otimes \left(e_{I}^{(n)}e_{J}^{(k)}\otimes u^{(n)}u^{\prime (k)}\right)
\end{eqnarray*}
 Expanding $e_{I}^{(n)}e_{J}^{(k)}=\sum _{L}m_{nI,kJ}^{(n+k)L}\, e_{L}^{(n+k)}$
(see (\ref{eq:m_A_basis}), (\ref{eq:m_A*_basis})) and equating both
parts we obtain\[
\sum _{KL}\left(\sum _{IJ}\overline{m_{nI,kJ}^{(n+k)K}}\, m_{nI,kJ}^{(n+k)L}\right)\, e^{(n+k)K}\otimes e_{L}^{(n+k)}=\sum _{K}e^{(n+k)K}\otimes e_{K}^{(n+k)}\qquad \qquad \forall n,k\]
This necessary condition for $(End_{\#}(A),\bullet ,\Delta )$
to be a bialgebra translates into
\begin{equation}
\sum _{IJ}\overline{m_{nI,kJ}^{(n+k)K}}\, m_{nI,kJ}^{(n+k)L}=\delta ^{KL}\qquad \qquad \forall n,k\label{eq:graded_sum_mm_condition}\end{equation}
It can be written as
\begin{equation}
m_{A}\left((D_{A}a)^{(n-k,k)}\right)=a\qquad \qquad \forall a=a^{(n)}\in A_{n},\quad 0\leq k\leq n\label{eq:homo_condition_graded}\end{equation}
where $(D_{A}a)^{(n-k,k)}\in A_{n-k}\otimes A_{k}$ is the term of
the coproduct having first (resp. second) factor of grading $n-k$ (resp. $k$).
The condition (\ref{eq:m_D_condition}) should therefore be satisfied  \emph{for
each term of definite grading} of the coproduct of $a$.

\subsubsection{The case of $Paths$}

The above equation (\ref{eq:homo_condition_graded}) is verified in
the case of the algebra $Paths$. This is easy to see%
\footnote{Here we are \emph{not} using the bialgebra structure on
$Paths$ mentioned in section \ref{sec:paths_on_graph}  coming
from the group-like comultiplication, but rather a coproduct $D_{A}$
obtained from the concatenation product $m_{A}$ via the use of a
scalar product as in \ref{sub:A_and_A*}.%
} as $D_{A}a$ ---being dual to $m_{A}$--- gives all the possible
{}``cuts'' of $a$. Taking $a$ to be an elementary path of
length (grading) $n$, $(D_{A}a)^{(n-k,k)}$ is simply the cut where
the second factor has length $k$. Concatenating back both factors
we re-obtain $a$ again. Thus we have a (graded) bialgebra structure on $End_{\#}(Paths)$.
What is less evident is that we have also the same property for  $End_{\#}(\mathcal{E})$, when
it is endowed with the {\sl graded} multiplication $\bullet$, as proven in this paper (sec.  
\ref{sec:weak_bialgebra_condition_proof}).

\end{document}